# Scaling Data-Driven Building Energy Modelling using Large Language Models

**Sunil Khadka**
*Student Member ASHRAE*

**Liang Zhang, PhD**
*Associate Member ASHRAE*

**ABSTRACT**

Building Management System (BMS) through a data-driven method always faces data and model scalability issues. We propose a methodology to tackle the scalability challenges associated with the development of data-driven models for BMS by using Large Language Models (LLMs). LLMs' code generation adaptability can enable broader adoption of BMS by "automating the automation," particularly the data handling and data-driven modeling processes. In this paper, we use LLMs to generate code that processes structured data from BMS and build data-driven models for BMS's specific requirements. This eliminates the need for manual data and model development, reducing the time, effort, and cost associated with this process. Our hypothesis is that LLMs can incorporate domain knowledge about data science and BMS into data processing and modeling, ensuring that the data-driven modeling is automated for specific requirements of different building types and control objectives, which also improves accuracy and scalability. We generate a prompt template following the framework of Machine Learning Operations so that the prompts are designed to systematically generate Python code for data-driven modeling. Our case study indicates that bi-sequential prompting under the prompt template can achieve a high success rate of code generation and code accuracy, and significantly reduce human labor costs.

## INTRODUCTION

Buildings contribute to approximately 30% of the world's total energy consumption and 26% of global greenhouse gas emissions (IEA 2023). Effective building operation and control strategies are crucial for optimizing energy usage (Costa, Keane et al. 2013). One way to achieve this is through a data-driven Building Management System (BMS) (Maddalena, Lian and Jones 2020) that uses machine learning and statistical modeling techniques to extract insights and develop predictive models to inform decision-making processes for building operations and control. These systems rely on various data sources, including weather data, occupancy patterns, energy consumption, and indoor environmental conditions (Amasyali and El-Gohary 2018). However, the widespread adoption of data-driven BMS faces scalability challenges. Each building has different characteristics and extensive data requirements, along with the need for customized data-driven models, which make modeling difficult for building operations. Additionally, the development and deployment of these systems often require significant manual effort and specialized expertise.

To address these challenges, LLMs (Vaswani, Shazeer et al. 2017) have emerged as a powerful technology that can generate human-like text, code, and data representations. LLM is a deep-neural network-based natural language processing model to understand and generate contextual and coherent outputs based on input prompts. LLMs achieve excellence by undergoing extensive training on diverse datasets (Brown, Mann et al. 2020), (Izacard, Lewis et al. 2022), that encompass a wide range of programming languages, code examples, technical guides, and discussion forums. This comprehensive training

**Sunil Khadka** is a graduate student in the Department of Civil and Architectural Engineering and Mechanics, University of Arizona, Tucson, Arizona. **Liang Zhang** is an assistant professor in the Department of Civil and Architectural Engineering and Mechanics, University of Arizona, Tucson, Arizona.

equips them with the ability to understand and replicate the syntax and meaning of multiple programming languages, making them highly adaptable for various coding tasks. LLMs utilize the transformer architecture (Vaswani, Shazeer et al. 2017), which effectively utilizes self-attention mechanisms(Vaswani, Shazeer et al. 2017). The self-attention mechanism is crucial in LLM models for code generation because it allows the model to focus and weigh the important parts of the input sequence, regardless of their position for generating coherent and contextually aligned outputs (Wong, Guo et al. 2023). LLMs can perform tasks better by fine-tuning, leading to more accurate and relevant code-generation outcomes (Liu, Zeng et al. 2023).

Prompt (Liu, Yuan et al. 2023) is a collection of instructions that are given to an LLM to program it by customizing, enhancing, or refining its capabilities. Its objective is to influence future interactions and output generated from an LLM by providing specific rules and a predetermined set of instructions for a discussion (White, Fu et al. 2023). This crucial process in which the input prompts given to an LLM are designed and refined to produce specific and desirable outputs is called Prompt Engineering (Schmitt 2024) (White, Fu et al. 2023). A key component in prompt engineering (Liu, Yuan et al. 2023) involved the development of well-structured prompt templates which include placeholders like {adjective} or {content}. These can be customized by users with specific inputs to generate the intended prompt. With well-engineered prompts, prompt engineering can help in the improvement of content generation, and accurate and relevant responses.

One of the key advantages of LLMs is their ability to generate code and automate various tasks with prompts, including data processing, analysis, and modeling (Chen, Tworek et al. 2021). This feature has great potential to solve the scalability issue in data-driven modeling development for building operations. However, only a few studies have been found in this field, such as the scope of LLM in renewable energy (Rane 2023), and energy load prediction (Zhang, Lu and Zhao 2024). Yet, the existing literature has not fully addressed how prompt engineering can enhance model scalability and efficiency in these applications. Hence, this paper proposes a framework that leverages the code generation capabilities of LLMs to address the scalability challenges associated with the development of data-driven models used in BMS.

## METHODOLOGY

The proposed methodology includes the development of prompt engineering templates, serving as a foundational bone structure, in an organized storyline to generate code successfully for data-driven modeling of building energy use. The prompt template is designed in accordance with MLOps workflow (AmazonWebServices 2019), to provide a standardized procedure of data-driven/machine-learning modeling.

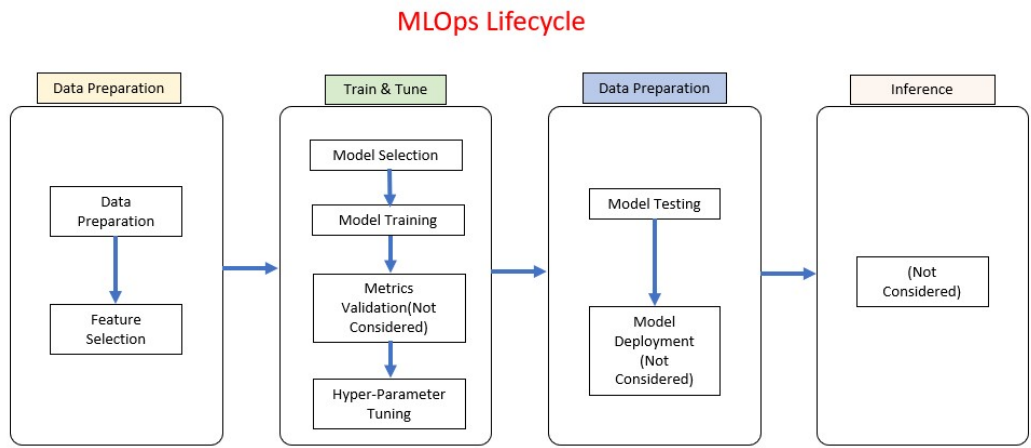

**Figure 1** MLOps lifecycle used for prompt template creation. (AmazonWebServices 2019)

The MLOps in the prompt template include 1) data preparation, 2) feature engineering, 3) model algorithm selection, 4) model training, 5) hyper-parameter tuning, and 5) model evaluation. The workflow does not include inference and model deployment steps as mentioned in Figure 1. This prompt template, serving as a foundational bone structure, aggregates all the components into an organized storyline to create the template of code generation for data-driven modeling of building energy use. The final prompt template following MLOps structure is presented in Table 1 serving as the reference prompt template for the research.

<div style="text-align:center">**Table 1. Proposed Prompt Template**</div>

| |
|---|
| Step 1 Data Preparation: Generate a Python code for a machine learning model using these CSV files: *{_ insert input data file name _}* for input data and *{_insert output data file name_}* for output data. The code should be adaptable for datasets with varying numbers of rows and columns and machine learning steps must be categorized in the following systematic steps. *{_insert model purpose = "This model is used to predict energy consumption one time step ahead"}* |
| Step 2 Feature Selection: Please generate a code to select important features from *{_ insert input data file name_ }* that impact target variables in *{ _ insert output data file name _}* using the best feature selection method in machine learning. The methods can be correlation analysis, recursive feature elimination, principal component analysis, wrapper method, or filter method. |
| Step 3 Data Splitting: Generate the code to design the dataset splitting strategy to divide the dataset into three distinct subsets: the training set, validation set, and test set. Begin by assessing the dataset characteristics, considering its size, potential class imbalances, and the complexity of the target machine learning problem. Considering the overall dataset size, establish a balanced partitioning scheme, such as *{%}* for training, *{%}* for validation, and *{%}* for testing. |
| Step 4 Model Selection: Please provide me with a snippet that can help me choose suitable regression models independently such as linear regression and random forest. |
| Step 5 Hyper Parameter Tuning: Please use optimizing model hyperparameters such as grid search or random search, as well as techniques for performing cross-validation to ensure the model's robustness for both linear regression and random forest models. |
| Step 6 Model Evaluation: After selecting the model, generate code to evaluate it on the test set using relevant metrics. For example: *{_input relevant metrics_}* |
| Step 7: Presentation of Machine Learning Model in Graph. Generate a code to create a detailed and visually appealing set of two scatter plot images side byside on a single canvas. The first scatter plot should represent a Linear Regression model's actual versus predicted values. The plot should be titled 'Linear Regression\nActual vs. Predicted' with labeled axes: 'Actual values' on the X axis and 'Predicted values' on the Yaxis. The second scatter plot, immediately to the right of the first, depicts the same concept for a Random Forest model. Title this plot 'Random Forest\nActual vs. Predicted' and label the axes similarly to the first plot. |

We proposed three methods for interacting with an LLM via prompt templates: 1) one-shot prompting, 2) step-wise sequential prompting, and 3) Bi-sequential prompting. One-shot prompting delivers all responses from the LLM at once through a single prompt, minimizing user-LLM exchanges and conserving time and effort. In contrast, step-wise sequential prompting breaks a complex query into smaller, sequential prompts, allowing for a more exploratory, step-by-step approach. Bi-sequential prompting, meanwhile, splits the entire prompt into two sequences for a structured, two-part exploration. Our evaluation will focus on the LLM's response quality, including prompt completion rate and code accuracy, and the interaction time cost, employing these prompting techniques.

## CASE STUDY

### Virtual Building Settings

The case study takes place in a DOE reference building (small office) with an area of 511m2 (5500sqft). This building was developed by the National Renewable Energy Laboratory and is simulated using EnergyPlus software for modelling. More details about the building can be found in the work done by (Deru, Field et al. 2011). The weather file for this modelling is USA_AZ_Tucson.Intl.AP.722740_TMY3.epw. The following is the variables of the simulation: zone temperature setpoint (ZTSP), zone air temperature (ZSF1_ZT), outdoor air dry-bulb temperature (OADT), direct solar radiation rate per area (SR_DIR), occupancy (OCC), and single zone cooling rate (Clg_Rate). The modeling is a regression to predict zone cooling rate (output), using the remaining variables (input). The generated data are saved in two csv files 'input_fx.csv' and 'output_fx.csv', which are generated after data processing which contains the input variables and output variables respectively for BMS.

### Large Language Model Settings

We choose ChatGpt-4, March 14, 2023 version, released by OpenAI (OpenAI 2023) as the LLM for the case study.

## RESULTS

Using the prompt template, we carried out the experiment among three prompting strategies. The performance is defined by two factors: 1) prompt completion rate: the success rate for LLM to generate all the code for each step without termination, 2) code accuracy: the success rate of the generated code to run error-free. Since our objective is to reduce human labor through these results, we planned to compare these results with manual coding. We assigned the same task to a machine learning beginner who has been learning machine learning for few months, to evaluate the time duration, performance outcomes for the same task.

**One-shot prompting strategy**

The strategy known as the one-shot prompting strategy has been specifically designed for the purpose of obtaining a response through the use of a single prompt. In order to evaluate the effectiveness of this strategy, the experiment was carried out a total of 35 times, resulting in a prompt completion rate of 71.42% (23 out of 35 attempts). The result is illustrated in Table 1.

**Table 2. One Shot Prompt and Its Response**
(Source: chat.openai.com/share/05c0ec3a-8239-4109-abec-34da8c5cb429)

| Prompt: |
|---|
| I have uploaded two files for a machine learning model i.e. 'input_fx.csv' and 'output_fx.csv'. Read the contents of the file properly. Imagine you are a machine learning engineer who has to generate a code for this task with the following steps. Step 1: Data Preparation Generate a Python code for a machine learning project using these CSV files: 'input_fx.csv' for input data and 'output_fx' for output data. The code should be adaptable for datasets with varying numbers of rows and columns and machine learning steps must be categorized in the following systematic steps. Step 2: Feature Selection Please generate a code to select important features from 'input_fx.csv' that impact target variables in 'output_fx.csv' using the best feature selection method in machine learning. The methods can be recursive feature elimination, correlation analysis, principal component analysis, wrapper method, or filter method. Step 3: Data Splitting Generate the code to design the dataset splitting strategy to divide the dataset into three distinct subsets: the training set, validation set, and test set. Begin by assessing the dataset characteristics, taking into account its size, potential class imbalances, and the complexity of the target machine learning problem. Considering the overall dataset size, establish a balanced partitioning scheme, such as 80% for training, 10% for validation, and 10% for testing. Step 4: Model Selection Please provide me with a snippet that can help me choose suitable regression models independently like linear regression, random forest. Step 5: Hyperparameter tuning Please use optimizing model hyperparameters such as grid search or random search, as well as techniques for performing cross-validation to ensure the model's robustness for both linear regression and random forest model. Step 6: Model Evaluation: Generate code to evaluate both models on the test set using relevant metrics (e.g. mean squared error). Also, Plot the graph to visualize results at the end. Step 7: Presentation of Machine Learning Model in Graph Generate a code to create a detailed and visually appealing set of two scatter plot images side-by-side on a single canvas. The first scatter plot should represent a Linear Regression model's actual versus predicted values. The plot should be titled 'Linear Regression\nActual vs. Predicted' with labeled axes: 'Actual values' on the X-axis and 'Predicted values' on the Y-axis. The second scatter plot, immediately to the right of the first, depicts the same concept for a Random Forest model. Title this plot 'Random Forest\nActual vs. Predicted' and label the axes similarly to the first plot. |

```
Response:
import pandas as pd
from sklearn.model_selection import train_test_split

# Load the datasets
input_data = pd.read_csv('/mnt/data/input_fx.csv')
output_data = pd.read_csv('/mnt/data/output_fx.csv')
print(input_data.head())
print(output_data.head())
data = pd.concat([input_data, output_data], axis=1)
data = data.fillna(data.mean())
from sklearn.feature_selection import RFE
from sklearn.linear_model import LinearRegression
X = data.iloc[:, :-1]  # all columns except the last one
y = data.iloc[:, -1]   # the last column

# Feature selection
model = LinearRegression()
rfe = RFE(model, n_features_to_select=5)  # Adjust based on your preference
fit = rfe.fit(X, y)

# Print out the selected features
selected_features = X.columns[fit.support_]
```

```
# Example for Linear Regression
param_grid_lr = {
    # Parameters for linear regression tuning
}
lr_grid_search = GridSearchCV(estimator=linear_model,
param_grid=param_grid_lr, cv=5)
lr_grid_search.fit(X_train, y_train)
# Example for Random Forest
param_grid_rf = {
    'n_estimators': [100, 200],
    'max_depth': [None, 10, 20],
}
rf_grid_search = GridSearchCV(estimator=random_forest_model,
param_grid=param_grid_rf, cv=5)
rf_grid_search.fit(X_train, y_train)
from sklearn.metrics import mean_squared_error
import matplotlib.pyplot as plt

# Evaluate models
lr_predictions = lr_grid_search.predict(X_test)
rf_predictions = rf_grid_search.predict(X_test)
# Plotting the results
plt.figure(figsize=(10, 5))
```

```
print("Selected Features: ", selected_features)          # Linear Regression plot
                                                          plt.subplot(1, 2, 1)
# Splitting the data                                      plt.scatter(y_test, lr_predictions)
X_train, X_temp, y_train, y_temp = train_test_split(X, y, plt.title('Linear Regression\nActual vs. Predicted')
test_size=0.2, random_state=42)                           plt.xlabel('Actual values')
X_val, X_test, y_val, y_test = train_test_split(X_temp, y_temp,  plt.ylabel('Predicted values')
test_size=0.5, random_state=42)
from sklearn.ensemble import RandomForestRegressor        # Random Forest plot
from sklearn.linear_model import LinearRegression         plt.subplot(1, 2, 2)
                                                          plt.scatter(y_test, rf_predictions)
# Initialize models                                       plt.title('Random Forest\nActual vs. Predicted')
linear_model = LinearRegression()                         plt.xlabel('Actual values')
random_forest_model = RandomForestRegressor()             plt.ylabel('Predicted values')
                                                          plt.tight_layout()
from sklearn.model_selection import GridSearchCV          plt.show()
```

The detailed results of LLM's experiments are summarized into three categories and detailed in Table 3. It was observed that approximately 28.57% (10/35) of errors occurred during the initial step due to the inability of the LLM to accurately read and incorporate input variables from CSV files. These errors can largely be attributed to the excessively large size of the chunks provided to the LLM, leading to the omission of crucial initial steps and an overall inability to effectively analyze the files. In terms of model accuracy, it was found to be 57.14% (20 out of 35 attempts). This raises substantial questions about the chunk size limit in one-shot prompt strategy that LLMs can accurately manage for successful code generation.

**Table 3. Results for One-Shot Prompting**

| S.N. | Scenario Description | Occurrence (Total Attempts: 35) | Source |
|---|---|---|---|
| 1 | Success: Code is generated successfully without any error. | 23 | chat.openai.com/share/05c0ec3a-8239-4109-abec-34da8c5cb429 |
| 2 | Few-Prompts Missed: One or two prompts are auto missed by LLM. Example in this case, few steps for linear regression are ignored. | 2 | chat.openai.com/share/107b8c8d-c0df-43da-8b8d-523b32b7a05e |
| 3 | Error: LLM due to computational error cannot read the files and give code without reading the file. Further-more, Chat-GPT4 computational error causes the system overhang and crash midway through code generation process. | 10 | chat.openai.com/share/bd479067-88a6-449c-bb05-0dece7b55f45 |

.

In terms of model accuracy, it was found to be 57.14% (20 out of 35 attempts), which is lower than the prompt completion rate. This raises substantial questions about the chunk size limit in one-shot prompt strategy that LLMs can accurately manage for successful code generation.

**Bi-Sequential prompting strategy**

In this strategy, we split the prompts into two sequential parts, termed bi-sequential prompting strategy. The first part included the first three steps, while the second part consisted of the remaining four steps. The outcomes achieved by implementing the sequential prompting approach surpassed those obtained from the one-shot prompting method, exhibiting a prompt completion rate of 100% and a code accuracy of 97.14% (34/35 attempts). The results are presented below in Table 4 along with a link to the experiment.

**Table 4.   Results for Bi-Sequential Prompting**
(Source: chat.openai.com/share/ecf3c827-659a-43e1-b8af-08b65f246bef)

| | |
|---|---|
| Prompt:<br>I have uploaded two files for a machine learning model i.e 'input_fx.csv' and 'output_fx.csv'. Read the contents of the file properly. Imagine you are a machine learning engineer who has to generate a code for this task with the following steps. Step 1: Data Preparation Generate a Python code for a machine learning project using these CSV files: 'input_fx.csv' for input data and 'output_fx.csv' for output data. The code should be adaptable for datasets with varying numbers of rows and columns and machine learning steps must be categorized in the following systematic steps. Step 2: Feature Selection Please generate a code to select important features from 'input_fx.csv' that impact target variables in 'output_fx.csv' using the best feature selection method in machine learning. The methods can be recursive feature elimination, correlation analysis, principal component analysis, wrapper method, or filter method. Step 3: Data Splitting Generate the code to design the dataset splitting strategy to divide the dataset into three distinct subsets: the training set, validation set, and test set. Begin by assessing the dataset characteristics, taking into account its size, potential class imbalances, and the complexity of the target machine learning problem. Considering the overall dataset size, establish a balanced partitioning scheme, such as 80% for training, 10% for validation, and 10% for testing. | |
| Response:<br>import pandas as pd<br>from sklearn.model_selection import train_test_split<br>from sklearn.feature_selection import SelectKBest, f_regression<br># Load data<br>input_fx = pd.read_csv('input_fx.csv')<br>output_fx = pd.read_csv('output_fx.csv')<br><br># Step 1: Data Preparation (Assuming data is already clean)<br><br># Step 2: Feature Selection<br># Merge input and output for correlation analysis<br>data = pd.concat([input_fx, output_fx], axis=1)<br># Use SelectKBest for feature selection<br>selector = SelectKBest(score_func=f_regression, k='all')<br>X_new = selector.fit_transform(input_fx, output_fx.squeeze()) | #Assuming output_fx is a single column<br>selected_features = input_fx.columns[selector.get_support()]<br><br># Step 3: Data Splitting<br>X = input_fx[selected_features]  # Only selected features<br>y = output_fx<br># Splitting the data into training (80%), validation (10%), and testing (10%)<br>X_train, X_temp, y_train, y_temp = train_test_split(X, y, test_size=0.2, random_state=42)<br>X_val, X_test, y_val, y_test = train_test_split(X_temp, y_temp, test_size=0.5, random_state=42)<br># Output the shapes of the datasets to confirm splitting<br>print(f"Training set: {X_train.shape}, Validation set: {X_val.shape}, Testing set: {X_test.shape}") |
| Prompt:<br>Step 4: Model Selection Please provide me with a snippet that can help me choose suitable regression models independently like linear regression, random forest. Step 5: Hyperparameter tuning Please use optimizing model hyperparameters such as grid search or random search, as well as techniques for performing cross-validation to ensure the model's robustness for both linear regression and random forest model. Step 6: Model Evaluation: Generate code to evaluate model on the test set using relevant metrics (e.g. mean squared error). Also, Plot the graph to visualise results at the end. Step 7: Presentation of Machine Learning Model in Graph Generate a code to create a detailed and visually appealing set of two scatter plot images side-by-side on a single canvas. The first scatter plot should represent a Linear Regression model's actual versus predicted values. The plot should be titled 'Linear Regression\nActual vs. Predicted' with labeled axes: 'Actual values' on the X-axis and 'Predicted values' on the Y-axis. The second scatter plot, immediately to the right of the first, depicts the same concept for a Random Forest model. Title this plot 'Random Forest\nActual vs. Predicted' and label the axes similarly to the first plot. | |
| Response:<br>from sklearn.linear_model import LinearRegression<br>from sklearn.ensemble import RandomForestRegressor<br>from sklearn.model_selection import GridSearchCV, cross_val_score<br>from sklearn.metrics import mean_squared_error<br>import numpy as np<br>import matplotlib.pyplot as plt<br><br># Initialize models<br>linear_model = LinearRegression()<br>random_forest_model = RandomForestRegressor(random_state=42)<br># Grid for Random Forest<br>rf_grid = {'n_estimators': [100, 200], | # Random Forest with best parameters<br>best_rf_model = rf_grid_search.best_estimator_<br>rf_predictions = best_rf_model.predict(X_test)<br><br># Evaluation<br>linear_mse = mean_squared_error(y_test, linear_predictions)<br>rf_mse = mean_squared_error(y_test, rf_predictions)<br>print(f"Linear Regression MSE: {linear_mse}")<br>print(f"Random Forest MSE: {rf_mse}")# Plotting the results<br>plt.figure(figsize=(14, 6))<br><br># Linear Regression plot<br>plt.subplot(1, 2, 1)<br>plt.scatter(y_test, linear_predictions, alpha=0.5)<br>plt.title('Linear Regression\nActual vs. Predicted') |

```
        'max_depth': [None, 10, 20],
        'min_samples_split': [2, 5],
        'min_samples_leaf': [1,
rf_grid_search = GridSearchCV(estimator=random_forest_model,
param_grid=rf_grid, cv=5, scoring='neg_mean_squared_error',
verbose=1)
rf_grid_search.fit(X_train, y_train.squeeze())
print(f"Best parameters for Random Forest:
{rf_grid_search.best_params_}")
# Linear Regression
linear_model.fit(X_train, y_train)
linear_predictions = linear_model.predict(X_test)

plt.xlabel('Actual values')
plt.ylabel('Predicted values')
plt.plot([y_test.min(), y_test.max()], [y_test.min(), y_test.max()],
'k--', lw=2)
# Random Forest plot
plt.subplot(1, 2, 2)
plt.scatter(y_test, rf_predictions, alpha=0.5, color='red')
plt.title('Random Forest\nActual vs. Predicted')
plt.xlabel('Actual values')
plt.ylabel('Predicted values')
plt.plot([y_test.min(), y_test.max()], [y_test.min(), y_test.max()],
'k--', lw=2)
plt.show()
```

**Step-wise sequential prompting strategy**

In this strategy, each step is performed sequentially, one at a time, leading to successful code generation with a prompt completion rate of 100%. However, the code accuracy was significantly lower at 60% (21/35). The stark contrast between the completion rate and code accuracy can be attributed to code repetition; subsequent prompts read the initial prompt instructions and repeat the same steps, causing the model to execute identical steps multiple times. chat.openai.com/share/0ccfb7a5-e3ac-4369-97bb-4139ca6b294d.

**Comparison with human coding**

The mean squared value is 0.437 and 0.0533 for linear regression and random forest in manual coding, whereas that by LLM is 0.313 and 0.047. The reduction in mean squared error for both linear regression (28.38%) and random forest (11.82%) when using LLM instead of manual coding indicates that LLM provide more accurate predictions, reflecting a significant improvement in model performance. The manual approach required over 2 days for the subject to achieve the desired efficient results, showcasing how LLM can reduce both time and human costs, with higher probability of accuracy.

**DISCUSSION**

We compare three prompting strategies: one-shot prompting, step-wise sequential prompting, and bi-sequential prompting. Bi-sequential prompting performs the best because its prompt completion rate is 100%, better than or equal to the other two (71.42% and 100%); the code accuracy is 97.14%, which is also better than the other two (57.14% and 60%).

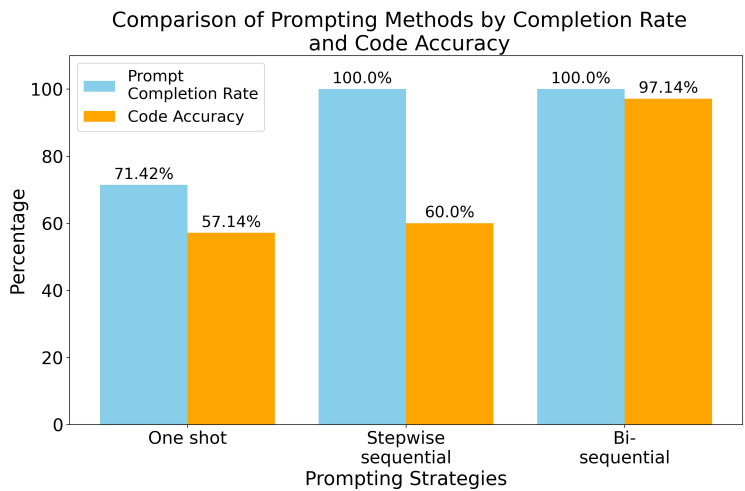

**Figure 2**   Bar chart representing the accuracies for three prompting experiments.

Table 5. Accuracy Comparison Among Prompting Strategies

| Accuracy | One-shot prompting | Step-wise sequential prompting | Bi-sequential prompting |
| --- | --- | --- | --- |
| Prompt completion rate | 71.42% | 100% | 100% |
| Code Accuracy | 57.14% | 60% | 97.14% |

This indicates the chunk size of prompts matters in the code generation. If the chunk size is too large (meaning we put everything into one prompt), the LLM will lose track of long instructions in the prompts; If the chunk size is too small, the LLM tends to repeat code as it revisits previous prompts; its premature prediction of future code steps based on its training on diverse datasets, attempting to anticipate the next step before it was actually requested. This inclination of LLM to forecast future steps and codes adversely affected the accuracy of the step-wise sequential results. To mitigate these challenges, the bi-sequential prompt strategy finds a balancing point of chunk size to achieve high accuracy in code generation tasks.

A beginner's manual coding approach for MLOps tasks like feature selection and hyperparameter tuning can be time-consuming due to their limited understanding and experience. In contrast, LLM have been trained on vast amounts of data, including code repositories and machine learning algorithms, giving them a comprehensive knowledge base. For example, in our case study, LLM considers the grid parameters like n_estimators, max_depth, min_samples_split, and min_samples_leaf, whereas a beginner takes time to select between the hyperparameters that might be good for this operation, leading to more time for project completion and higher probability of less accurate results. Additionally, the task of replicating the same process for different buildings poses a challenge due to the unique characteristics of each building, requiring adjustments or complete rewrites of the code.

## CONCLUSION

The paper proposes a prompt template following the framework of Machine Learning Operations so that the prompts are designed to systematically generate Python code for data-driven modeling. The prompt template's consistency and reliability in the case study shows its effectiveness and potential for a wider range of data-driven modeling in buildings. Our further investigation into various prompting strategies, including one-shot prompting, step-wise sequential prompting, and bi-sequential prompting. We find bi-sequential prompting delivered the best performance in prompt completion rate and code accuracy. The study also highlights the critical importance of prompt length and sequence. Excessively detailed prompts can overwhelm LLMs, leading to suboptimal solutions and repetitive code generation as in the case of step-wise sequential prompting. Also, when compared with the manual coding, it reduces code development time from hours/days to seconds.

In terms of limitations, since we have used Chat-GPT-4, March 14, 2023 version as LLM for case study, the performance of the proposed prompt template running in other LLMs is unknown. Second, the specific needs or customized requirements are not fully incorporated within our standard template. We only incorporate one placeholder for modeling purpose, but in the future, we can add more details about building types, system types, and other information that might impact the selection of features and machine learning algorithms. Moreover, the use of multi-agent workflow is not considered in this paper. By enabling parallel processing, these systems significantly accelerate problem-solving by allowing multiple agents to work on different aspects of a problem simultaneously. This approach may speed up the trial-error cycle but also optimizes resource allocation by dynamically assigning tasks based on each agent's capabilities and workload. Additionally, our study is limited by the absence of comparable research in the field of prompt engineering in other disciplines, highlighting this as future exploration work.

## NOMENCLATURE

BMS = Building Management System  
LLM = Large Language Model